\documentclass[preprint]{elsarticle}
\usepackage[english]{babel}
\usepackage{sidecap}
\usepackage[utf8]{inputenc}
\usepackage{amsfonts}
\usepackage{subfigure}
\usepackage{amsmath}
\usepackage{graphicx}
\usepackage{colortbl}
\usepackage{color}
\usepackage{xcolor}
\usepackage{placeins}
\usepackage{url}
\usepackage{slashbox}
\usepackage{vmargin}
\usepackage{natbib}
\usepackage{amssymb}
\usepackage{amsthm}
\usepackage{multicol}
\usepackage{rotating}
\usepackage[english,ruled,vlined]{algorithm2e}

\journal{Neurocomputing}

\begin{document}

\begin{frontmatter}

\title{Analysis of Professional Trajectories  using Disconnected Self-Organizing Maps}
\author[ifsttar]{Etienne Côme}
\author[paris1]{Marie Cottrell}
\author[paris12]{Patrice Gaubert}
\address[ifsttar]{Universit\'{e} Paris-Est, IFSTTAR, GRETTIA, F-93166 Noisy-Le-Grand, France}
\address[paris1]{Laboratoire SAMM, Universit\'e Paris 1 Panthéon-Sorbonne\\
90 rue de Tolbiac, F-75634 Paris Cedex 13, France}
\address[paris12]{ERUDITE, Universit\'e Paris Est Cr\'eteil,\\
61, avenue du G\'eneral De Gaulle,  94010 Cr\'eteil, France }
\date{\today}



\begin{abstract}
In this paper we address an important economic question. Is there, as mainstream economic theory asserts it, an homogeneous labor market with mechanisms which govern supply and demand for work, producing an equilibrium with its remarkable properties?  Using the Panel Study of Income Dynamics (PSID) collected on the period 1984-2003, we study the situations of American workers with respect to employment. The data include all heads of household (men or women) as well as the partners who are on the labor market, working or not. They are extracted from the complete survey and we compute a few relevant features which characterize the worker's situations.

To perform this analysis, we suggest using a Self-Organizing Map (SOM, Kohonen algorithm) with specific structure  based on planar graphs, with disconnected components (called D-SOM), especially interesting for clustering. We compare the results to those obtained with  a classical SOM grid and a star-shaped map (called SOS). Each component of D-SOM takes the form of a string and corresponds to an organized cluster. 

From this clustering, we study the trajectories of the individuals among the classes by using the transition probability matrices for each period and the corresponding stationary distributions. 

As a matter of fact, we find clear evidence of heterogeneous parts, each one with high homogeneity, representing situations well identified in terms of activity and wage levels  and in degree of stability in the workplace.
These results and their interpretation in economic terms contribute to the debate about flexibility which is commonly seen as a way to obtain a better level of equilibrium on the labor market.

\end{abstract}

\end{frontmatter}

\section{Introduction}

The aim of this study is to identify and to analyze the succession of situations occupied by workers on a modern labor market (1984-2003), it is an extended version of \cite{wsom2012}.

Basically the dominant economic theory (neo-classical sense) is based on the concept of the labor market where supplies (individuals) and demands (firms) meet. An equilibrium price (salary) makes the adequacy of supply and demand \cite{cahuc2004}. 
This theory defines mechanisms explaining labor supply by the wage level and predicting the stability of the relation between a firm and a worker and its evolution over time (a career). Unfortunately, these mechanisms are not observed in most real situations.

This is the pure neo-classical model. To get closer to the real economy, many developments have been made in the representation of the behavior of economic agents, in particular with the theory of job search taking into account different types of imperfections (incomplete information, the presence of institutions, regulation of relations between firms and employees...), see for example chapter 39 in \cite{handbook1999}. But the basic hypothesis is unchanged: a single market whose functioning is flawed with the constraints and inefficiencies caused by the actual conditions. The result is closer to the real economy, but the deeper understanding of the phenomena that affect this sector of the economy in the last 30 years and their dynamics in the changing conditions of this period cannot be properly identified.

Instead of completing the pure neo-classical market model by a set of constraints more or less complex, the idea of this work is to show that the market is not homogeneous, it is the assembly of parts whose main characteristics are very different: that is the meaning of the assertion that the labor market is heterogeneous as mentioned in  \cite{ilo1} and \cite{ilo2}. To find evidence of this heterogeneity, we construct a classification of the labor market observed over twenty years: this should identify the specific characteristics of each component and, secondly, permit to observe the situation of employees, over time, in these specific markets.

Our contribution consists in the identification and characterization of each class essentially using the variables used for classification (as well as some qualitative variables, subject to a satisfactory quality of this information.) A result to be expected is the dynamic view of trajectories of employees between these classes which can be obtained by observing the transition matrices between classes.

Let us  identify the diversity of situations in terms of activity.  
A ``situation''  is defined by a combination of quantitative variables as  shown in these two examples: 
\begin{itemize}
\item  full time job for the whole year, high wages, seniority in the same job;

\item  precarious conditions, wages lower than the average, part-time jobs, short-term contracts, on-call jobs, holding  of a second job.
\end{itemize}

On the basis of individual characteristics, we construct a classification of  situations observed every 2 years on a specific labor market, the US labor market  over two consecutive periods of nine (1984-1992) and eleven (1993-2003) years respectively. So for each individual, we can observe the sequence of the classes he belongs to. That is what we call \textit{professional trajectories}.

We need to study the temporal changes to answer some important questions linked to the evolution of the macroeconomic environment. Recall that in 1992, the end of Reaganomics and the beginning of Clinton period lead to a global reduction of unemployment during the rest of the decade. And this change leads to ask several questions. For instance what are the real changes at the individual level after 1992? More generally, what is the impact of a reduction of unemployment associated with ``new forms of employment'' . And also are there different conclusions if we observe the specific situation of women on the labor market?


This article follows another paper \cite{come10} but contains necessary material (and possibly redundant) to be self-contained. It is organized as follows:  Section 2 presents the data and the notations used throughout the paper. The methodology and the global architecture of the proposed procedure are described in Section 3. Section 4  discusses how to choose the more efficient topology for the map. In sections 5 and 6, the classes are analyzed from an economic point of view. Finally section 7  presents the transitions from one class to another, according to the period and gender. Section 8 is devoted to a discussion of recent articles and to a conclusion which summarizes the main results.

\section{The Data: first period (1984, 86, 88, 90, 92) and second period (93, 95, 97, 99, 2001, 03) }
\label{sec:data}

We use the PSID (Panel Study of Income Dynamics)\footnote{Available online at \url{http://psidonline.isr.umich.edu/}}, dividing the observations in two periods in order to meet two objectives: on one hand to observe a number of workers large enough to obtain statistical indicators representative of the whole population and on the other hand to keep only individuals present all along each period to identify trajectories.

We create a sample for each period (1984-1992, 1993-2003). By looking at descriptive statistics for the quantitative variables for each period, we can assume that both periods have the same rough characteristics. So we can make the classifications with all the observations together.

In the PSID, we select households for which the head (man or woman) is present in the household every year of the period and we do it separately for each period. The administrative rule is that if there is a male in the household, he is the head, if not the head is a woman. Fortunately quite the same variables concerning the activity on the labor market are available for the wife/partner of the head, if there is one. Retrieving this information, we constitute a set of individuals (3965  in period 1, and 3607 in period 2) observed every two years in each period, with a proportion of women close to that observed in the whole population.

An observation consists of a couple (year, individual). Each one is described by 8 quantitative variables and 2 qualitative variables. See Table \ref{tab:varlist} for the list of variables and their meaning.

\begin{table}[h]
\begin{center}
\begin{tabular}{p{1cm}p{2cm}p{5.5cm}p{3cm}p{2cm}}\hline
&\textbf{Name}& \textbf{Description}	& \textbf{Min-Max}	& \textbf{Type}		 \\\hline\hline
  		& nbhtrav &	Number of worked hours per week  &  0-112 &  Quant       \\
        & nbstrav  &	Number of worked weeks       &   0-52 &  Quant       \\
        & nbschom  &	Number of unemployed weeks   &   0-52 &  Quant       \\       
  & nbsret   &	Number of weeks out of the labor market & 0-52 &  Quant      \\
        & salhor   &	Hourly wages in real dollars      & 0-83.85 & Quant       \\
        & nbex     &	Number of extra jobs     &   0-5     &  Quant       \\    
    & hortex  & Number of hours worked in extra jobs & 0-1664 &  Quant       \\   
    & anctrav  & Seniority in  current job in months &  0-780 &  Quant       \\
\hline
     & gender      &	Gender        &  2 modalities       	&  Qual        \\   
     & age  & Age group ($<$30, 30-45, $>$45) & 3 modalities &  Qual        \\   
\hline    
\end{tabular}
\end{center}
\caption{\label{tab:varlist} Variable names, description and type, for PSID dataset.}
\end{table}

The pre-processing consist of removing observations with clearly inconsistent values such as a number of week per year greater than 52. After this filtering 41467 observations constitute our working database. Observed current wages are converted in real dollars using the Price Index of PIB in 1992 (first period) or 2003 (second period). Eventually, the 8 quantitative variables were centered and reduced to standardize the order of magnitude. We can compute the correlation matrix of these variables, displayed in Table~\ref{tab:corr}.

\begin{table}[h]
\begin{center}
\begin{tabular}{l|rrrrrrrrr}
	&\begin{sideways}\textit{nbhtrav}\end{sideways}&\begin{sideways}\textit{nbstrav}\end{sideways}&\begin{sideways}\textit{nbschom}\end{sideways}&\begin{sideways}\textit{nbsret}\end{sideways}&\begin{sideways}\textit{salhor}\end{sideways}&\begin{sideways}\textit{nbex}\end{sideways}&\begin{sideways}\textit{hortex}\end{sideways}&\begin{sideways}\textit{anctrav}\end{sideways}&\\
\hline		
\textit{nbhtrav}     &	1	&	0.72	&	-0.04	&	-0.14	&	0.36	&	0.05	&	0.01	&	0.23		&\\
\textit{nbstrav}     &	0.72	&	1	&	-0.23	&	-0.30	&	0.38	&	0.06	&	0.01	&	0.30		&\\
\textit{nbschom}   &	-0.04	&	-0.23	&	1	&	0.02	&	-0.09	&	-0.01	&	-0.01	&	-0.11		&\\
\textit{nbsret}       &	-0.14	&	-0.30	&	0.02	&	1	&	-0.10	&	-0.04	&	-0.04	&	-0.12		&\\
\textit{salhor}       &	0.36	&	0.38	&	-0.09	&	-0.10	&	1	&	0.07	&	0.05	&	0.31		&\\
\textit{nbex} 	       &	0.05	&	0.06	&	-0.01	&	-0.04	&	0.07	&	1	&	0.72	&	0.00		&\\
\textit{hortex}     &	0.01	&	0.01	&	-0.01	&	-0.04	&	0.05	&	0.72	&	1	&	-0.01		&\\
\textit{anctrav}    &	0.23	&	0.30	&	-0.11	&	-0.12	&	0.31	&	0.00	&	-0.01	&	1		&\\
\end{tabular}
\end{center}
\caption{\label{tab:corr} Correlation matrix of the quantitative variables.}
\end{table}

We observe that variables \textit{Number of worked hours per week} (nbhtrav), \textit{Number of worked weeks} (nbstrav), \textit{Hourly wages in dollars} (salhor) and \textit{Seniority in  current work in months} (anctrav) are strongly positively correlated, and that they are opposite to \textit{Number of unemployed weeks} (nbschom) and \textit{Number of weeks out of the labor market} (nbsret). The variables related to \textit{extrajobs} are not correlated with the others.

\section{SOM, Disconnected Self-Organizing Maps (D-SOM), Self Organizing Star (SOS)}

\subsection{The Kohonen algorithm (SOM)}

In its classical presentation \cite{Kohonen95, cottrell98}, the SOM algorithm is an iterative algorithm, which iterates the two following steps over training patterns $\mathbf{x}_j$ for computing the set of code-vectors $\mathbf{m}_i, i\in\{1,\hdots,K\}$ which define the map:

\begin{itemize}
\item \textit{Competitive step,} this step aims at finding the best matching unit (BMU)  for sample $\mathbf{x}_j$ : 

\begin{equation}
c = \arg\min_{i\in\{1,\hdots,K\}}||\mathbf{x}_j-\mathbf{m}_i||.
\end{equation}

\item \textit{Cooperative step,} this step aims at moving the code-vectors of the BMU and of  its neighbors (on the map), closer to the training pattern:

\begin{equation}
\mathbf{m}_{i}(t+1)=\mathbf{m}_{i}(t)+\alpha(t)h_{ci}(t)\left[\mathbf{x}_j-\mathbf{m}_i(t)\right],
\end{equation}

with $t$ the time step, $\alpha(t)$ the learning rate of the algorithm and $h_{ci}(t)$ the neighborhood function between units $c$ and $i$ at time $t$.
\end{itemize}

Several neighborhood functions are commonly used such as $h_{ci}(t)=\mathbf{1}_{(d_{ci}<\sigma_t)}$ or $h_{ci}(t)=\exp(-d_{ci}^2/2\sigma_t^2)$. All of them depend on a radius $\sigma_t$ which is classically decreasing during the learning process. These neighborhood functions also depend on $d_{ci}$: the distance between units $c$ and $i$, which is determined by the map topology.

During the cooperative step, the coordinates of the best matching code-vector 
$\mathbf{m}_c$ are updated in order to move it closer to the training pattern. Other code-vectors $\mathbf{m}_i$ are also moved towards the training pattern according to their distance $d_{ci}$ from the BMU defined by the lattice. Closer code-vectors from the BMU are more affected than the others. The two steps are iterated over the dataset during several epochs until convergence. Thanks to the cooperative steps, self-organization is reached at the end of the algorithm. 

One can see that the distance between units  in the map plays a key role in the self-organization property of the algorithm; modifications of this distance will have an impact on the results of the algorithm. Therefore, the lattice structure can be a way to incorporate prior information concerning the dataset topology into the dimensionality reduction process.  In the context of classical SOM, one assumes that the dataset topology can effectively be represented by a grid or a straight line but other hypotheses can be interesting and can advantageously be investigated. 

It has been already noted that graph theory can be used to define this distance \cite{barsi03,pakkanen06}. In this case, map units are the nodes of a graph, and distances between them are defined as the minimum number of edges needed to reach one node starting from the other, that is the so-called \textit{shortest path distance}. We therefore propose to modify the SOM algorithm only by taking as input an adjacency matrix  \footnote{For two units $i$ and $j$, entry $(i,j)$ of the adjacency matrix is 1 if there exists an edge between $i$ and $j$, and 0 if not.} (see~\cite{godsil2001}), which specifies the graph topology that the user desires. All undirected graphs can theoretically be used, but a specific class is of interest: the class of planar graph, because such graphs can easily be represented in a 2-dimensional setting allowing us to supply the SOM with visualization tools, as seen in the next section.    



\subsection{New topologies}

An interesting choice is defined by a map which is composed of several disconnected one-dimensional strings. Each string will contain data which are similar at a rough level and that are displayed in  an ordered disposition. 


This topology has a special interest: when  the map consists of not connected parts, the  "cooperation" step of the algorithm only concerns the units which belong to the same component as the winning unit. The competition step is not modified, so that the algorithm meets a double goal :

\begin{enumerate}
\item to group the observations into macro-classes corresponding to the different disconnected components of the graph;
\item to organize the units inside the macro-classes.
\end{enumerate} 

\begin{figure}[h!]
\begin{center}
\begin{tabular}{ccc}
\includegraphics[scale=0.4]{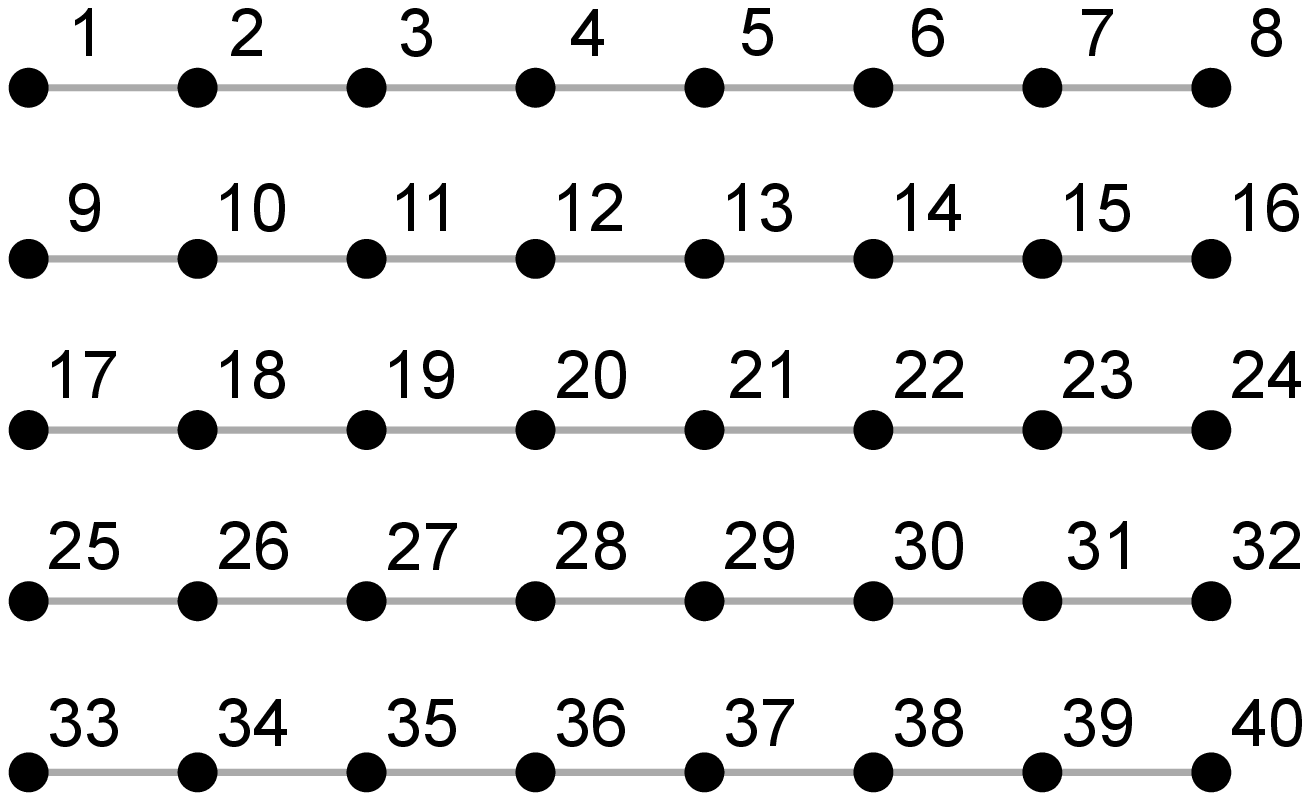}&&\includegraphics[scale=0.4]{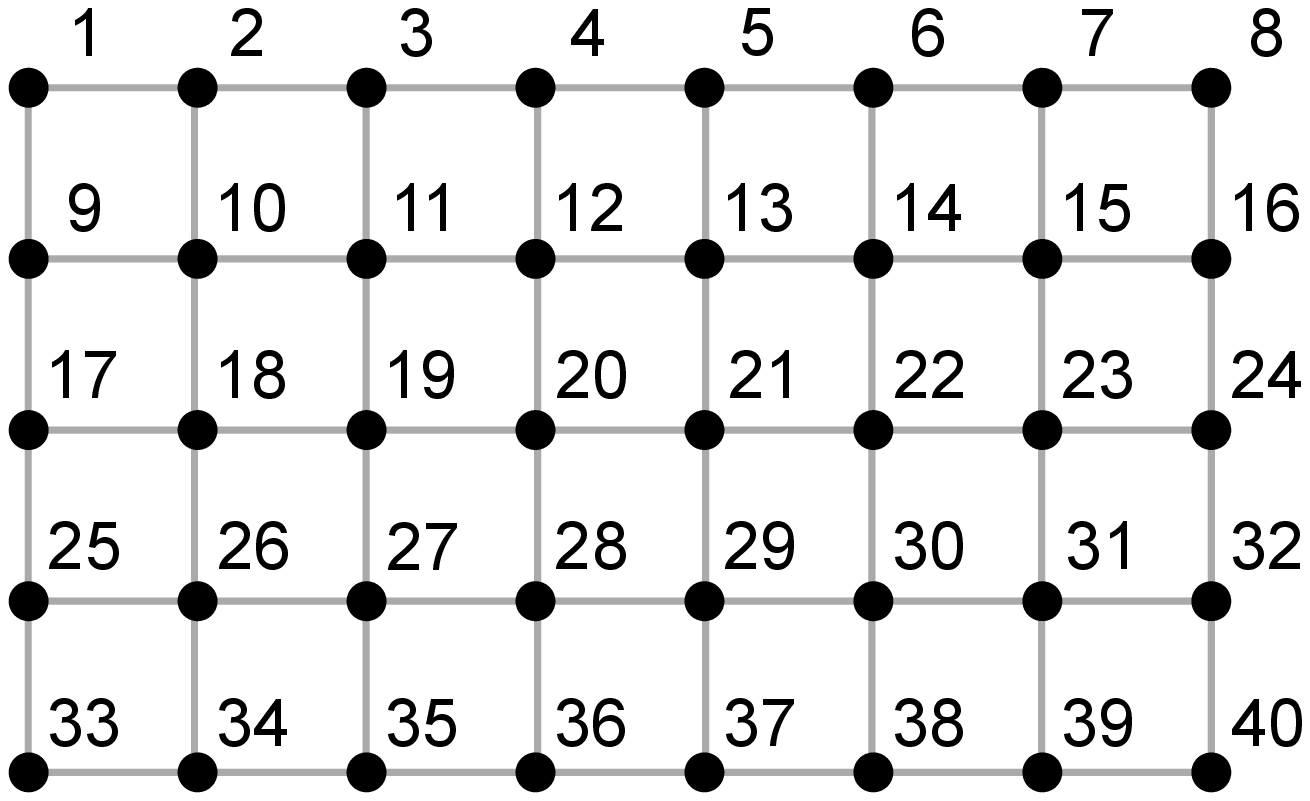}\\
(a) && (b)
\end{tabular}
\caption{\label{fig:topol}(a) Two-dimensional representation of a disconnected map with 5 strings of 8 units and (b) representation of classical (5 by 8) grid map.}
\end{center}
\end{figure}

Figure \ref{fig:topol} shows an example of a disconnected neighborhood structure that we define here and of a classical grid neighborhood. In the disconnected case, for example, $d_{13,21}=+\infty$ and $d_{26,31}=5$, while in the classical grid case, $d_{13,21}=1$ and $d_{26,37}=4$.

In conclusion, by imposing a limitation of the cooperation which only acts inside the macro-classes and by keeping a competition between all units, this algorithm allows us to obtain a given number of macro-classes which are themselves self-organized. We call this topology the D-SOM.

\begin{figure}[h!]
\begin{center}
\includegraphics[scale=0.3]{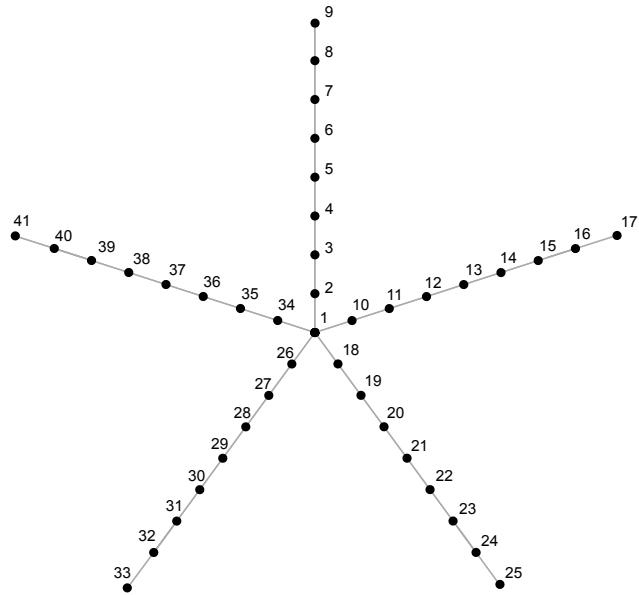}
\caption{\label{fig:topolstar}Two-dimensional representation of a star-shaped map with 5 rays of 8 units.}
\end{center}
\end{figure}

Another interesting choice is a star-shaped graph as shown in Figure \ref{fig:topolstar}. This graph has a clear natural center from which different arms or rays grow. Such graph can easily be interpreted  by users if different rays correspond to different well identified classes while the center gather the ``normal''   patterns. The organization  takes place on each  ray and distances to the center  describe the characteristics of the patterns in an ordered way. These graphs can be characterized by their number of rays and by the length of the rays. SOM using such a topology will be called Self-Organizing Stars (SOS) (as defined in Come et al. (2010) \cite{come10}).
   
Figure \ref{fig:topolstar} shows an example of star-shaped neighborhood structure. In this case, for example $d_{3,20}=5$.\\

There exist other methods to obtain well-separated classes, see \cite{resta11} for example.  But our approach is different since we do not look for building an adjacency matrix between the code-vectors by repeating many runs of the SOM algorithm. Contrarily, we impose an a priori adjacency matrix which defines star-shaped classes or non-connected classes.

It is also possible to use U-matrix visualization as in \cite{ultsch1990} or \cite{vesanto2000a}, or direct clustering of code-vectors as in \cite{vesanto2000b} to define the macro-classes. In the following we achieve an  Hierarchical Ascending Classification of the code-vectors in the classical grid case, to group the classes into a previously fixed number of macro-classes, see \cite{cottrell98}. \\

Both kinds of topology (D-SOM, SOS) are well adapted to the analysis of labor market segmentation, since one looks for a segmentation into macro-classes  well discriminated,  split into organized classes. In a general case, the question of the choice of the number of macro-classes is guided by \textit{a priori} argument if it exists. In our case, we choose 5 macro-classes which is the best choice to get contrasting and well identified situations. In fact, in the literature, the authors generally choose 4 segments, (for example, see two recent studies on real markets, the French  and the German labor markets, \cite{ilo1}, \cite{ilo2}). But we have to add a fifth segment  specific to the US economy,  which corresponds to the regular practice of two or more jobs (it is a fairly common practice in the US economy, while it is rare in France and Germany). This is the reason for choosing five macro-classes and no other numbers that the economic model probably could not explain.

Let us now describe the results that we get using these three topologies for the PSID data.

\section{Comparison of the maps, choice of a topology}

We use the Kohonen algorithm for three different topologies : the classical one on a (5 by 8) grid, a D-SOM with 5 strings of 8 units, a SOS with 5 rays of 8 units. The total numbers of classes are almost the same (40, 40 and 41 units). The data include 41467 couples (year, individual) represented by a 8-vector composed of the 8 quantitative variables.

The number of initial classes (micro-level) is determined by the number of available observations: descriptive statistics constructed at this level  must be calculated with a sufficient number of observations in each of these classes. With approximately 40,000 observations, dividing each macro-class in 8 units gives 40 classes with 1000 observations by class on average.

\begin{figure}[h!]
\begin{center}
\begin{tabular}{c}
\includegraphics[scale=0.62]{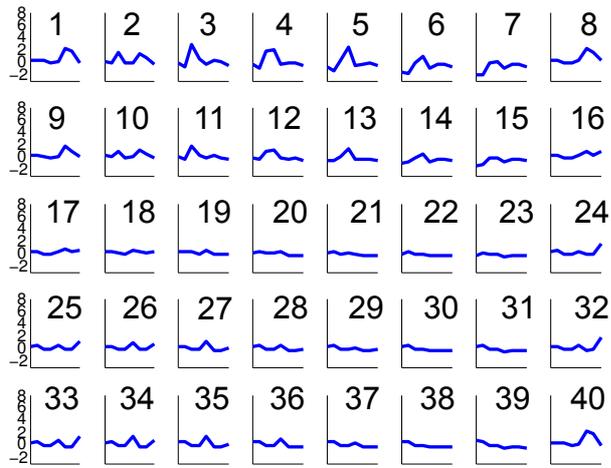}\\
(a)\\
\includegraphics[scale=0.62]{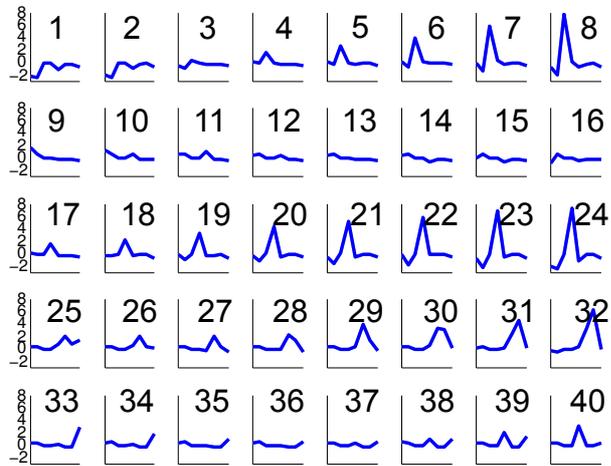}\\
(b)\\
\includegraphics[scale=0.62]{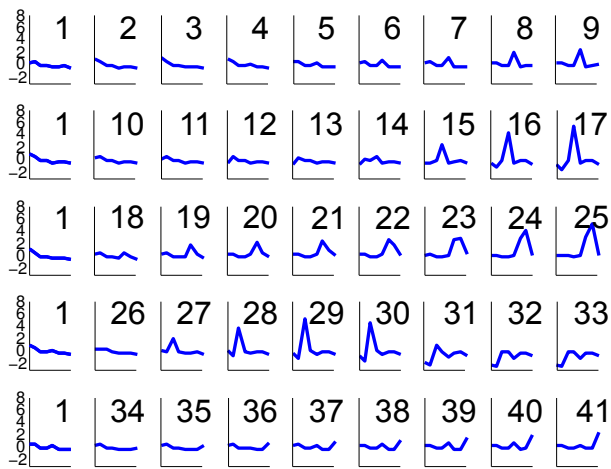}\\
(c)\\
\end{tabular}
\caption{\label{fig:maps} (a) SOM map (classical grid neighborhood) codebook representation, (b) D-SOM map codebook representation, (c) SOS map codebook representation. For each codebook its coordinates in the feature space are depicted with features ordered as in Table~\ref{tab:varlist}. For each subplot, in abscissa we find the features number and in ordinate the standardized values of the features for the codebook.}
\end{center}
\end{figure}

Figures  \ref{fig:maps} (a), (b) and (c) show the code-vectors for each map: the 8 components of each code-vector are displayed according to the order defined in Table \ref{tab:varlist}. They are well organized. The SOM map is organized in all directions, while the others are organized inside each string.

From a quantitative point of view, a \textit{first indicator} to compare the three maps is the \textit{quantization error}, which is a measure of quality of the clustering. We consider the within sum of squares as:

\begin{equation}
SC_{within}=\sum_{x}\|x-\mathbf{m}_{c(x)}\|^2,
\end{equation}

where
\begin{equation}
c(x)=\arg\min_{i\in\{1,\hdots,K\}}||\mathbf{x}-\mathbf{m}_i||.
\end{equation}

This is simply the sum of the squared distances  between each pattern $x$ and the code-vector of its BMU, in the pattern space.

Then we define the total sum of squares as
\begin{equation}
 SC_{total}= \sum_{x}\|x-\bar{x}\|^2.
\end{equation}

So we can define the \textit{relative quantization error} as:

\begin{equation}
  RQE=\frac{SC_{within}}{SC_{total}}.
\end{equation}

The smaller the relative quantization error, the better the classification. \\

A \textit{second indicator} is the ratio between the sum of squares extended to neighbor code-vectors and the total sum of squares.
If we note (as in \cite{ritter92})

\begin{equation}
  SC_{extended}= \sum_{x}\sum_{k \in \mathcal{V}(c(x))} \frac{1}{|\mathcal{V}(c(x))|}||x-m_k||^2,
\end{equation}

where $\mathcal{V}(c(x))$ is the set of neighbors of $c(x)$, as defined by the adjacency matrix of the graph, we can compute the \textit{relative extended quantization error}:

\begin{equation}
 RQE_{ext}=\frac{SC_{extended}}{SC_{total}}.
\end{equation}

A small value of $RQE_{ext}$ indicates a good organization, since it implies that neighbor code-vectors on the map are close in the pattern space.

For comparison, the three possible solutions namely SOM, D-SOM and SOS were fitted to the PSID data with the same procedure and parameters (linear decreasing learning rate and Gaussian neighborhood function one run per method), the corresponding results are given in Table~\ref{tab:Quanti}.

Table \ref{tab:Quanti} shows that D-SOM gets better quantization than the others at the unit level.  This is because the map is less constrained, the adaptation algorithm finds a better minimum for $RQE$. As to the relative extended quantization  error $RQE_{ext}$, the results of D-SOM and SOS are close and better than the SOM case results. Furthermore, D-SOM and SOS allow us to get well-contrasted and easy-to-interpret macro-classes. As our main goal is to build robust macro-classes, we decide to consider only these two topologies.

\begin{table}[h]
\begin{center}
\begin{tabular}{l|rr}
		&	\textbf{$RQE$}		&\textbf{$RQE_{ext}$}	\\\hline		
SOM	   	 &	22.23\%    			&	40.37\%	\\
D-SOM		&	12.79\%	 		&	22.01\%       	\\
SOS	   	 &	16.79\%	 		&	22.36\%      	\\
\end{tabular}
\end{center}
\caption{\label{tab:Quanti} Relative quantization error and relative extended quantization error for the three topologies in \%.}
\end{table}

At the unit level the D-SOM topology achieves therefore the best result on this particular dataset and such a topology seems therefore 
better fitted to such a case.\\

 The same type of quality measures can be computed at the macro-classes level, as a \textit{third indicator}. To this end we define:
\begin{equation}
SC_{macro}=\sum_{x}\|x-\mathbf{M}_{b(x)}\|^2,
\end{equation}
where $b(x)$ give the macro-class number of $x$ such as $b(x)=s, s\in \{1,\hdots,S\}$ with $S$ the number of macro classes in the map and $\mathbf{M}_s$ is the empirical mean of macro-class $s$ members. Normalizing this quantity by the total sum of square as previously give a normalized quality measure:
\begin{equation}
  RQE_{macro}=\frac{SC_{macro}}{SC_{total}}.
\end{equation}

\begin{figure}[h!]
\begin{center}
\includegraphics[scale=0.45]{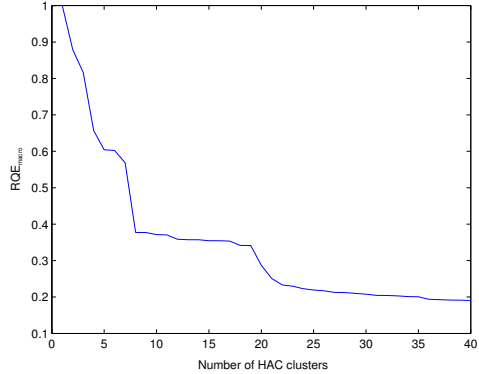}
\caption{\label{fig:cah}Evolution of $RQE_{macro}$ with respect to the number of macro-classes for SOM+HAC.}
\end{center}
\end{figure}

This quantity can be easily computed for the D-SOM topology and also for the SOS topology if each ray of the star is associated to a macro-class plus one macro-class for the center of the star. Since the classical SOM topology based on rectangular grid did not define such macro-classes, one has to use an additional step to build them using K-means \cite{vesanto2000b} or Hierarchical Ascending Clustering (HAC) \cite{cottrell98}. We used the later here to build the macro-classes in the classical SOM map. Using such an approach, the evolution of $RQE_{macro}$ with respect to the number of macro-classes for the SOM map can be computed and is depicted in Figure~\ref{fig:cah}. So we are able to do the comparison of the \textit{relative quantization errors at the macro-class level} for the three maps with 5 macro-classes. The results are presented in Table~\ref{tab:QuantiMacro}.

\begin{table}[h]
\begin{center}
\begin{tabular}{l|rr}
		&	\textbf{$RQE_{macro}$}	\\\hline		
SOM+HAC	   	 &	60.4\%    		\\
D-SOM		&	47.5\%	      	\\
SOS	   	 &	55.7\%	  	\\
\end{tabular}
\end{center}
\caption{\label{tab:QuantiMacro} Relative quantization error at the macro-classes level for the three topologies in \%.}
\end{table}

This quality measure leads to the same conclusion as for the previous, D-SOM performs better on the PSID dataset and the differences with the two other approaches is clearer. As expected the results are poorer than at the unit level since the description of the data is coarser. All these elements lead us to choose the D-SOM topology with five macro-classes as a reference to study the PSID dataset in the rest of the paper.\\

Even if D-SOM seems to be the best choice, we can go deeper in the comparison between SOS and D-SOM in analyzing the classifications \textit{from an economic point of view}. The best one is the one offering the clearest interpretation of a specific labor market during almost twenty years, with the influence of changing economic policies and changing economic environment. The latent global interpretation is that this labor market is not a homogeneous market, but a set of sub-markets well differentiated in terms of level of activity, wages and seniority; the links between them, for instance the worker's ability to move from one segment to another one, are important points to understand the economic system. 

\subsection{Analysis of the macro-classes resulting from D-SOM algorithm}
\label{classesDSOM}

This topology is well adapted, applied to the whole sample, if the basic hypothesis ``the labor market is constituted of 5 segments well differentiated independently from the period of observation''  is true. If this was false, we could see that some macro-classes correspond to the first period and the others to the second one: this is not what we have found. See Table~\ref{tab:period1byclass}. All macro-classes are almost equally divided between the two periods.

\begin{table}[h]
\begin{center}
\begin{tabular}{l|rrrrrrr}
              &   Class 1    & Class 2 & Class 3 & Class 4  & Class 5 & Sample \\
 \hline
\textbf{Whole sample}		 &6051		&15597	&1664		&5043		&13112	&41467&\\
\textbf{Period 1} (\%)		& 61.4		 &47.3	& 58.4          	&47.0		&41.1	   	&47.8&\\
\textbf{Period 2} (\%)		& 38.6		 &52.7	& 41.6          	&53.0		&58.9	   	&52.2&\\
\end{tabular}
\end{center}
\caption{\label{tab:period1byclass} Share of each period in each class.}
\end{table}

The proportions are approximately the same for both periods in each class, as we see in Table~\ref{tab:period1byclass}.\\

\begin{table}[h!]
\begin{center}
\begin{tabular}{l|rrrrrrr}
 	&Class 1 	&Class 2 	&Class 3 	&Class 4   	&Class 5 	&Sample&\\ 
&6051	&  15597	&  1664  &  5043   & 13112	&    41467&\\ 

\hline
\textbf{nbhtrav} 	&12.34          &\textbf{42.42}	 &28.33	&40.24          	&41.90	            &37.04&\\
\textbf{nbstrav}	&9.63	         &\textbf{49.18}	 &21.67	&46.04	          	&48.01	            &41.55&\\
\textbf{nbschom}	&\textbf{6.87}	         &0.11	 &0.73	            &0.47	           	&0.10	            &1.16&\\
\textbf{nbsret}           &0.42	         &0.12	 &\textbf{30.09}           &0.36	          	&0.14	            &1.40&\\
\textbf{salhor}	 &3.37	         &11.92    &7.57	            &14.39	          	&\textbf{18.61}             &12.91&\\
\textbf{nbex}		&0.02	        &0.00	&0.04	            &\textbf{1.20}		&0.00		&0.15&\\
\textbf{hortex}	&4.70	       &0.05	&6.22	           &\textbf{458.44}	&0.98	           &57.02&\\
\textbf{anctrav}	&11.19       &35.16	&17.47	           &93.77            &\textbf{202.68}	&91.05&\\
\end{tabular}
\end{center}
\caption{\label{tab:meanbyclassDSOM} Mean values for each variable by the 5 D-SOM macro-classes and for the whole sample; the  figures in bold are the maximum values for each variable, the figures below class names are the class sizes.}
\end{table}

So we may summarize the main results of the classification  in Table~\ref{tab:meanbyclassDSOM}, where each column shows a macro-class: the eight variables used by the algorithm describe the similarities and differences between the segments from a quantitative point of view (means).

\begin{itemize}
\item \textbf{Class 1}: low activity measured by hours of labor by week and number of weeks of labor per year and significant periods of unemployment (close to 7 weeks); 

\item \textbf{Class 2}: high activity but for a wage lower than the whole average and a short seniority; this may be identified as a set of jobs with a great flexibility;

\item \textbf{Class 3}: mainly defined by a part-time activity over the year with more than half  the year out of the labor market and a very short seniority\,\footnote{Presumably, unemployment and other situations out of activity are not correctly declared by the individuals: in this macro-class there is an average of 7 weeks unemployed, less than one week temporarily out of the market and close to 10 weeks at work, letting 34 weeks of a year unexplained.};

\item \textbf{Class 4}: constituted with people having more than one job at the same time; the obtained wage is then greater than the average;

\item \textbf{Class 5}: constituted of the ``good jobs'', with high activity, wages approximately 50\% greater than the average and a great stability with quite 17 years in the same place.
\end{itemize}

\subsection{Analysis of the macro-classes obtained using SOS algorithm}

Comparing D-SOM and SOS algorithms, a first difference comes from the structure  of the map. The SOS structure is very convenient if the phenomenon under study  contains  a reference situation or ``\textit{normal}''  situation (the center of the star) and  several shifts or gaps of one or several variables  which are  represented on the different rays of the star. It could be a good representation if the labor market was conceived as a pure mechanism that could be observed in the real economic system (what is not true due to institutional and social constraints). 

From the computational point of view, the results are very similar to that obtained with D-SOM, although there is a sixth macro-class (the center of the star).

We calculate the arithmetic means of the eight variables and  can emphasize the contrasts between the macro-classes, (the five rays and the center). In Table~\ref{tab:meanbyclassSOS}, let C1-SOS, C2-SOS, ..., C6-SOS be the classes obtained using SOS algorithm. Class C6-SOS is the center of the star.

\begin{table}[h]
\begin{center}
\begin{tabular}{l|rrrrrrrr}
 	          &C1-SOS 	&C2-SOS 	&C3-SOS  	&C4-SOS 	&C5-SOS 	&C6-SOS & Sample &\\
 	          & 10177 &7210     &4913	&6146	&12447	&574	&41467&\\
\hline
\textbf{nbhtrav} 	&\textbf{48.12}	           &31.31	            &40.32	             &13.04	&41.56	             &43.08& 37.04&\\
\textbf{nbstrav}	&\textbf{48.64}	          &42.78	            &46.03	             &10.69	&48.19             &48.66& 41.55 &\\
\textbf{nbschom}	&0.10	          &0.21	            &0.47	             &\textbf{6.89}	            &0.08	             &0.02& 1.16 &\\
\textbf{nbsret}	 &0.15	          &\textbf{6.99}	            &0.35	             &0.43	            &0.13	             &0.13& 1.40 &\\
\textbf{salhor}	&\textbf{19.96}		&7.35		&14.27		&3.36		&14.71		&9.45& 12.91 &\\
\textbf{nbex}		&0.01		&0.01		&\textbf{1.21}		&0.02		&0.01		&0.00& 0.15 &\\
\textbf{hortex}	&1.29		&1.44		&\textbf{466.91}	&4.81		&1.39		&0.00& 57.02 &\\
\textbf{anctrav}	&52.34		&24.11		&86.04		&11.08		&\textbf{205.39}	&37.92& 91.05 &\\
\end{tabular}
\end{center}
\caption{\label{tab:meanbyclassSOS}  Characteristics of the 6 macro-classes; the  figures in bold are the maximum values for each variable, the figures below class names are the class sizes.}
\end{table}

Three macro-classes are very close to their equivalent in the D-SOM classification  (C2-SOS, C3-SOS and C4-SOS respectively for Class 3, Class 4 and Class 1 in the D-SOM classification): out of the market temporarily, more than one job at the same time, low activity with some period of complete unemployment.
The last two (the sixth being treated separately) show situations with a great level of activity and wages that are very high (class C1-SOS) or slightly above the global mean (class C5-SOS).

We encounter some problems in interpretation: in opposition to the first classification, the class C1-SOS associates high wages  and short seniority,  while C5-SOS associates a relatively moderate wages with more than 17 years of seniority. As a matter of fact this opposition was observed inside  macro-class Class 5 of good jobs (D-SOM classification) between the units: at one end of the string are found the jobs with high wages and low seniority and at the other end the exact opposite situation. This has been interpreted as a trade-off between these two positive characteristics: you have to move on the market to obtain better careers with significant higher wages. 

\subsection{Crossing the classifications}
Finally if we cross both classifications at the individual level, we can show more precisely what the problem is. See Table~\ref{tab:SOSbyDSOM}:

\begin{table}[h]
\begin{center}
\begin{tabular}{l|r|r|r|r|r|r|rrr}
	&	C1-SOS          &	C2-SOS	&C3-SOS	&C4-SOS	&C5-SOS	&C6-SOS	&Total	&	\\
\hline
Class 1	&	49	&	79	&	2	&	5900	&	21	&	0	&	6051	&	\\
Class 2	&	7874	&	5519	&	3	&	216	&	1416	&	569	&	15597	&	\\
Class 3	&	35	&	1608	&	2	&	11	&	3	&	5	&	1664	&	\\
Class 4	&	13	&	3	&	4905	&	5	&	117	&	0	&	5043	&	\\
Class 5	&	2206	&	1	&	1	&	14	&	10890	&	0	&	13112	&	\\
\hline
Total	&	10177	&	7210	&	4913	&	6146	&	12447	&	574	&	41467	&	\\
\end{tabular}
\end{center}
\caption{\label{tab:SOSbyDSOM}  Cross-tabulation of the SOS classification (in column) and DSOM classification (in row).}
\end{table}

This table presents how a macro-class obtained using one algorithm (D-SOM if one reads the table row by row) is split in several parts, each one corresponding to a macro-class produced by the other one (SOS) and whose frequencies are the figures read on the same line.
If both algorithms produce very similar classification each class obtained with the first algorithm is mainly observed in another class from the second one: most of the observations figuring as the total of a row are concentrated in one cell. The number identifying a class being attributed randomly by the computer, most of the time it will not be situated on the diagonal.
This is the case, in table 8, for classes D-SOM 1, 3 and 4:
\begin{itemize}
\item	the 6051 observations of D-SOM class 1 are mainly (5900) in SOS class 4 and residual numbers in the five other ones (6 classes are obtained with SOS);
\item	on the 1664 observations grouped in the D-SOM class 3, 1608 belong to the SOS class 2;
\item	it is quite the same with the 5043 observations of the D-SOM class 4: 4905 are in the SOS class 3.
\end{itemize}
The main difference between the results of the algorithms in question here is to be found in the contents of the classes 2 and 5 (D-SOM):
\begin{itemize}
\item	the main part (50.5\%) of D-SOM class 2 is also in the SOS class 1 with the same characteristics (high activity, wages higher than the average of the class);
\item	35.4 \% are in restricted activity (in hours per week or in number of weeks);
\item	9.1 \% of D-SOM class 2 belong to the SOS class 5 (longer seniority);
\item	the D-SOM class 5 is split in two parts: 16.8 \% belonging to the SOS class 1 (those with higher wages and short seniority) and the main part (83.8 \%) found in the SOS class 5 (very high seniority and wages above the average).
\end{itemize}

This difference has a major importance because the clear differentiation between a class with precarious jobs, weak wages and short seniority and what we called the class of good jobs is not visible in the classification produced by SOS. Whereas, the D-SOM algorithm led to a satisfactory interpretation of a characteristic of the class of good jobs: faster career progression is achieved by a change of job and therefore lower seniority where wages are higher.
The relationship between two key variables to build the classification is not visible in the result obtained with SOS and is what ultimately determines the choice of D-SOM. So this implies that the main problem observed is due to this exchange between the two classes of high activity: individuals with relatively low wages combined with high seniority have been transferred to the other ray.

\section{Visualization of the 5 macro-classes in D-SOM}

The comparison between the three topologies lead us to only consider the D-SOM map and  we will concentrate the rest of the analysis on it. One can find the description of the 5 classes in Section \ref{classesDSOM}.

The results of D-SOM can be displayed in several ways, and we present some of them in this section.

\begin{figure}[h!]
\begin{center}
\includegraphics[scale=0.8]{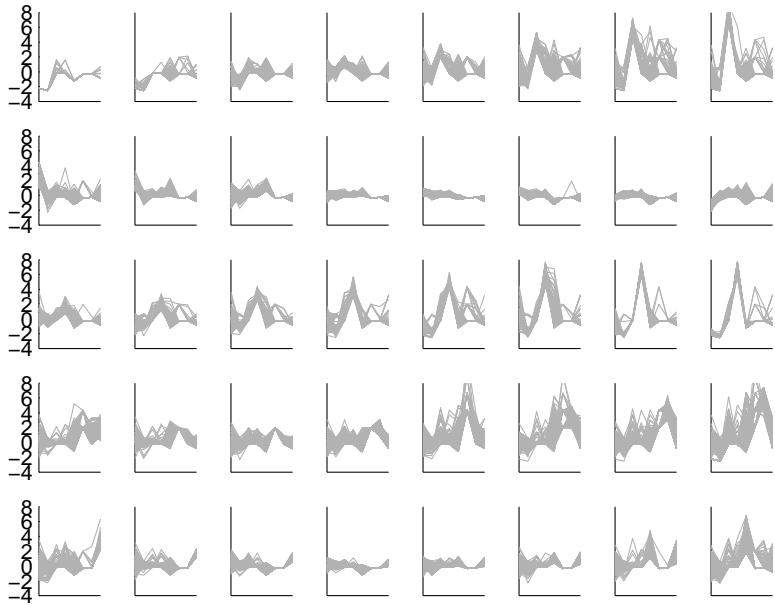}
\caption{\label{fig:map1} Contents of the classes  in the D-SOM map. For each subplot, in abscissa we find the features number and in ordinate the standardized values of the features, all members of the classes are represented by a line. }
\end{center}
\end{figure}

Figure \ref{fig:map1} shows the 41467 couples (year, individual) classified into 40 classes, and 5 disconnected macro-classes (the rows of the figure).  In each class, we draw  the stacking of individual lines obtained by connecting individual values (standardized) for the eight variables used for classification. It is a visual tool used to verify the homogeneity of the classes.  As each class is very homogeneous, this figure 
\ref{fig:map1} is very close to the figure \ref{fig:maps}-c, which represents the code-vectors.

\begin{figure}[htpb]
\begin{center}
\includegraphics[scale=0.8]{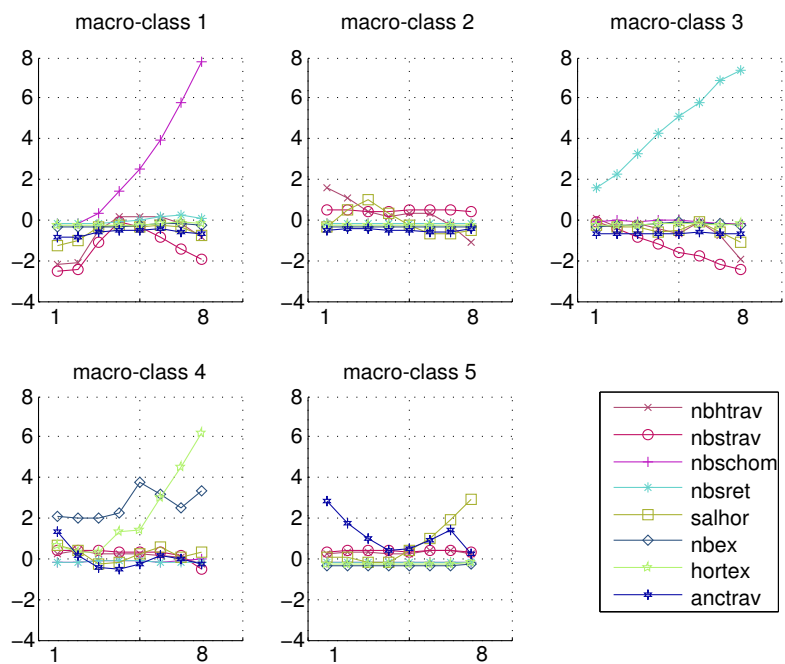}
\caption{\label{fig:map3} Multivariate profiles of the different macro-classes. For each subplot, in abscissa we find the unit number (inside the macro-class) and in ordinate the standardized values of the features for the codebook of each unit.}
\end{center}
\end{figure}

 Figure \ref{fig:map3} contains five subplots which present the evolution of the code-vectors along a macro-class from unit one to unit eight. All the variables are centered and reduced and are drawn on the same scale $[-4,\, 8]$. The major characteristic of each class can be seen by observing the noticeable variation of one (or two) of the variables used for classification, increasing or decreasing when going from the first to the last unit. 
 
More precisely, Figure \ref{fig:map3} shows that:
 \begin{itemize}
\item   Class 1 has a specific growth of the number of weeks of unemployment;
\item   Class 2 shows very weak salary and seniority while the number of hours of work is decreasing from unit 1 to unit 8;
\item   Class 3 is characterized by the number of weeks out of the market;
\item   Class 4 shows the importance of the two variables measuring the practice of two jobs or more;
\item   Class 5 combines a very high seniority and wages very superior to the sample average, but from the first to the last unit of this class, they change in exactly opposite directions.
\end{itemize}

\begin{figure}[htpb]
\begin{center}
\includegraphics[scale=0.8]{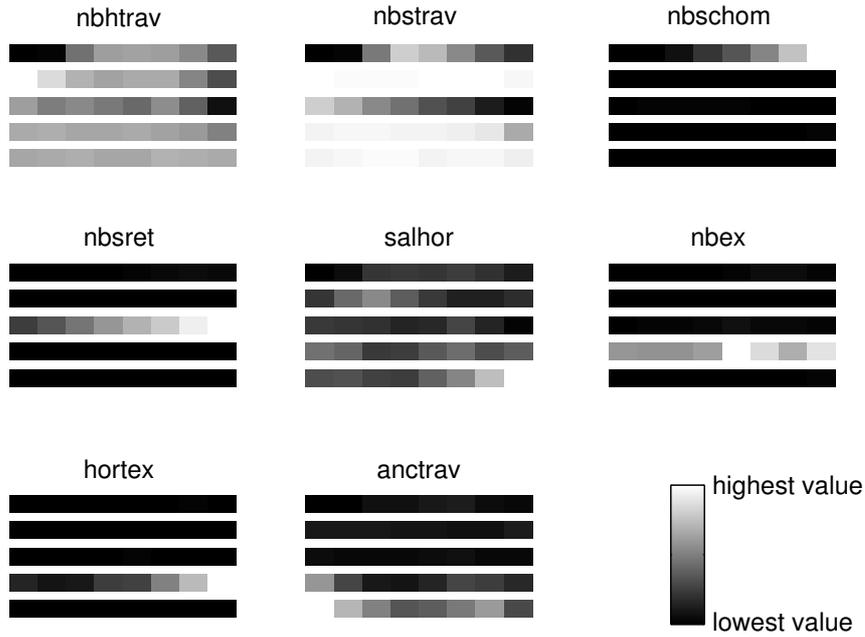}
\caption{\label{fig:map4}  8 variables on the whole D-SOM map, with 5 macro-classes of 8 units each one}
\end{center}
\end{figure}

Figure \ref{fig:map4} presents the 8 variables on the whole D-SOM map, with 5 macro-classes of 8 units each one. It is a  representation which is the  dual one of Figure \ref{fig:map3}: for each variable used in the classification, we see the extent to which a class is strongly influenced by it. For each variable, the 5 macro-classes are represented as stacked rows, showing the mean values computed at the unit level, using a color code from black for lowest value until white for the highest. It is a visual tool to compare the five classes through the importance of each variable.

All these representations confirm the descriptions of the 5 macro-classes we did in previous section.

\section{Crossing the classification obtained by D-SOM with exogeneous qualitative variables}

\subsection{Two periods}

We have already indicated that the two periods were very similar (see Table~\ref{tab:period1byclass}), but some changes must be reported (see Table~\ref{tab:distribyperiod}).

\begin{table}[h]
\begin{center}
\begin{tabular}{c|c|c}
Classes	&1984-92	&1993-2003	\\
\hline
Class 1		&18.75		&10.78		\\
Class 2		&37.22		&37.98		\\
Class 3		&4.90		&3.20		\\
Class 4		&11.95		&12.35		\\
Class 5		&27.18		&35.69		\\
\hline
Sample      &100        &100        \\
\end{tabular}
\end{center}
\caption{\label{tab:distribyperiod} Distribution of the 5 classes for both periods (in \%).}
\end{table}

\begin{itemize}
\item Significant reduction in the proportion of fully unemployed or employed a small part of the year (class 1), 18.75 to 10.78\%
\item Increase of the same order from the good jobs (Class 5) 27.18 to 35.69\%.
\end{itemize}
This is consistent with the overall finding made by all observers of the American macro-economy by comparing the late 80s and the 90s: the increased flexibility of the labor market results in a reduction in the overall unemployment and in an increase in various forms of employment.

\begin{table}[h!]
\begin{center}
\begin{tabular}{l|rrrrrr}\hline
	&	Class 1 &	Class 2 & Class 3  & Class 4 	&Class 5  	&	\\
\hline
Period 1 	&	 &	 &   &  	&  	&	\\
1984-92 	&	3717	&	7378	&	971	&	2370	& 5389	&	\\
\hline
Period 2	&	 &	 &   &  	&  	&	\\

1993-2003	&	2334	&	8219	&	693	&	2673	&	7723	&	\\
\hline

\textbf{nbhtrav}	&	13.53	&	\textbf{41.99}	&	27.03	&	40.77	&	41.76	&	\\
	&	10.44	&	\textbf{42.80}	&	30.15	&	39.76	&	42.00	&	\\
\hline
\textbf{nbstrav}	&	10.20	&	\textbf{48.69}	&	18.57	&	47.27	&	47.21	&	\\
	&	8.73	&	\textbf{49.62}	&	26.01	&	44.95	&	48.57	&	\\
\hline
\textbf{nbschom}	&	\textbf{7.65}	&	0.14	&	0.88	&	0.42	&	0.12	&	\\
	&	\textbf{5.61}	&	0.08	&	0.52	&	0.50	&	0.09	&	\\
\hline
\textbf{nbsret}	&	0.35	&	0.06	&	\textbf{31.73}	&	0.15	&	0.02	&	\\
	&	0.52	&	0.17	&	\textbf{27.79}	&	0.53	&	0.22	&	\\
\hline
\textbf{salhor}	&	3.44	&	10.74	&	6.43	&	11.75	&	\textbf{15.02}	&	\\
	&	3.24	&	12.99	&	9.18	&	16.73	&	\textbf{21.11}	&	\\
\hline
\textbf{nbex}	&	0.02 &	0.00	&	0.03& \textbf{1.13}	&	0.00	&	\\
	&	0.02	  &	0.00	&	0.06	&	\textbf{1.27}	&	0.01	&	\\
\hline
\textbf{hortex}	&	4.58	&	0.10	&	4.45	&	\textbf{344.76}	&	0.22	&	\\
&	4.90		&	0.01	&	8.69	&	\textbf{559.24}	&	1.51	&	\\
\hline
\textbf{anctrav}	&	11.97	&	35.06	&	15.33	&	84.87	&	\textbf{184.47}	&	\\
&	9.95	&	35.26	&	20.47	&	101.66 		&\textbf{215.39}	&	\\
\hline
\end{tabular}
\end{center}
\caption{\label{tab:meanperiod2} Mean values for each variable by macro-class, for periods 1 and 2 (one above the other), the  figures in bold are the maximum values for each row, the figures below the class names are the class sizes.}
\end{table}

According to Table~\ref{tab:meanperiod2} other categories vary slightly from one period to the other (the number of hours worked per week and number of weeks worked in the year, for example) for three active classes (Classes 2, 4 and 5). You still have to observe the growth of real wages for the class of good jobs (from 15.02 dollars per labor hour to  21.11 \$ / h), while for the class of precarious jobs hourly wage ranges from 10.74 to  12.99 \$ / h.

The resulting classes can be better defined by comparing the distribution of the sample according to the terms of age and gender.
\subsection{By gender}

According to table \ref{tab:sexbyclassbyperiod} for the entire observation period women are relatively more likely to be unemployed (more than 1 / 5) than men (1 / 13) and in return men are more often in precarious employment or perform two jobs simultaneously.
In the second period, the percentage of women in stable jobs is growing.

\begin{table}[h!]
\begin{center}
\begin{tabular}{l|r|r|r|r|r}
 By gender          &   Class 1    & Class 2 & Class 3 & Class 4  & Class 5 \\
    \hline
    	&Low &	Low skill	&Temporary 	&Two	&Good\\  
    	&   activity          & precarious   &Withdrawal& jobs            &jobs         \\
    	 \hline
Whole Sample 1 & & & & & \\
\textbf{Men}		&	7.64	&	40.71	&	1.64	&	15.29	&	34.70	\\
\textbf{Women}		&	22.32	&	34.17	&	6.65	&	8.68	&	28.19	\\ 
 \hline
Period 1 & & & & & \\
\textbf{Men}		&8.80	&	40.71	&	1.50	&	16.15	&	32.77	\\
\textbf{Women}		&	29.22	&	33.51	&	8.50	&	7.51	&	21.26	\\
\hline
Period 2 & & & & & \\
\textbf{Men}	&	6.56	&	40.72	&	1.77	&	14.54	&	16.13	\\
\textbf{Women}		&	15.69	&	34.79	&	4.86	&	9.81	&	34.85	\\
\end{tabular}
\end{center}
\caption{\label{tab:sexbyclassbyperiod} Structure in percentage by gender of the macro-classes for the whole sample and by period.}
\end{table}


\subsection{Age groups} 

The analysis of the results with respect to the age groups are given in Table~\ref{tab:agebyclass}. Most of the under than 30 are concentrated in two classes, Class 2 (48.37\%) and Class 1 (19.34\%), while the middle-aged (30 to 45) individuals are more present in the class of precarious activities (Class 2) and more than 45 predominate in Class 5. This is not surprising, but it reinforces the credibility of the obtained classification.

\begin{table}[h!]
\begin{center}
\begin{tabular}{l|r|r|r|r|r}
    By age          &   Class 1    & Class 2 & Class 3 & Class 4  & Class 5 \\
    \hline
    	&Low &	Low skill	&Temporary 	&Two	&Good\\ 
    	&   activity          & precarious   &Withdrawal& jobs            &jobs         \\
 \hline
\textbf{less than 30}, (8.66 \%)	&	19.34	&	48.37	&	8.44	&	11.54	&	12.32	\\
\textbf{from 30 to 45}, (60.64 \%)	&	12.81	&	39.06	&	3.97	&	12.80	&	31.36	\\
\textbf{more than 45}, (30.70 \%)	&	16.78	&	31.72	&	2.84	&	11.08	&	37.58	\\
\end{tabular}
\end{center}
\caption{\label{tab:agebyclass} Structure by age group of the macro-classes.}
\end{table}

Other qualitative variables such as skill and branch of work could be crossed with the obtained classification  in order to better describe the heterogeneity of subsets that are the real labor market. The quality of this information in PSID has deteriorated significantly in period 2 (inconsistencies between variables, updates not made from one period to another, notably in case of loss or change of job...). This degradation makes it very uncertain interpretation of crossings obtained with subsets of the market.
We choose not to present these results.

\section{Transitions between D-SOM macro-classes, empirical and limit distributions.}
\label{sec:markov}
In order to study the trajectories followed by individuals over each period, we consider that the successive situations are observations of a Markov chain.\\

Let us recall some elements of finite Markov chain theory 
(see for example \cite{hoel1987} or \cite{billing1979}). 

If $\{1, \ldots, K \}$ 
is the set of possibles states for a discrete process $X(t)$ (here $K=5$ and $X(t)$ is the class number of an observation at time $t$), the transition matrix $\Pi$ is a $K \times K$-matrix, with 

\begin{equation}
\Pi(i,j) = \mathbb{P}(X(t+1) =j\,|\, X(t)=i).
\end{equation}

Note that $0 \leq \Pi(i,j) \leq 1, \forall i, \forall j $ and that 
$\sum_j Pi(i,j)=1, \forall i$. Each row of the transition matrix sum to 1, and it is the probability distribution of the next state, conditionally to a starting position. One assumes that $\Pi(i,j)$ does not depend on $t$. Then such a discrete process $X(t)$ is a finite Markov chain, defined by its transition matrix $\Pi$.

We estimate the transitions probabilities $\Pi(i,j)$ by computing the empirical frequencies to be in Class $j$ at the next time, belonging in Class $i$ at present  \footnote{The data are observed each two years}. See in Table \ref{tab:transitionperiod}  the matrices  for both periods.
																							
\begin{table}[h]
\begin{center}
\begin{tabular}{l|rrrrr}
Period 1 	&	Class 1 &	Class 2 &	Class 3  &	Class 4  & Class 5  	\\
1984-92 	&	3717	&	7378	&	971	&	2370	&   5389		\\
\hline
Class 1	&	\textbf{56.51}	&	24.44	&	11.,73	&	4.11	&	3.69	\\
Class 2	&	8.44	&	\textbf{65.46}&	3.62	&	7.98	&	14.51	\\
Class 3	&	30.19	&	\textbf{43.01}	&	18.00	&	7.19	&	1.61	\\
Class 4	&	3.74	&	28.14	&	2.29	&	\textbf{50.76}	&	15.08	\\
Class 5	&	3.75	&	7.92	&	1.21	&	5.92	&	\textbf{81.21}	\\
\hline
Period 2	&	Class 1  &	Class 2  &	Class 3 &	Class 4 &	Class 5  \\
1993-2003	&	2334	&	8219	&	693	&	2673	&	7723	\\
\hline
Class 1	&	\textbf{57.67}	&	26.77		&	8.35	&	3.97		&	3.23	\\
Class 2	&	5.08		&	\textbf{67.66}	&	2.15	&	9.55		&	15.57	\\
Class 3	&	19.85		&	\textbf{52.73}&	13.18	&	7.51		&	6.74	\\
Class 4	&	2.79		&	29.44		&	1.23	&	\textbf{46.81}	&	19.74	\\
Class 5	&	2.34		&	13.98		&	0.96	&	6.47		&	\textbf{76.26}	\\
\end{tabular}
\end{center}
\caption{\label{tab:transitionperiod} Transition matrices, period 1 and period 2,  the figures below the class names are the class sizes, other values are expressed as percentages, values in bold are maxima of the row.}
\end{table}

The study of situations on the labor market can be realized in a dynamic sense, since the successive positions of each individual have been observed and used to construct the classification. 


The most visible result is that the major part of a class has not moved between year $t$ and year $t+2$: this is observed since the diagonal entries are the maximum of the line and even greater than 50, except for Class 3 and Class 4 in period 2.


Some global results can be pointed:
\begin{itemize}
\item  the transition from unemployment is mainly  towards unemployment itself and towards precarious situations of  Class 2 in a lower proportion;
\item  the transition from macro-class 2 towards  ``good jobs''  is slightly greater than the contrary (moves from Class 5 towards  Class 2);
\item  the most stable class over both periods is Class 5, the one with  ``goods jobs'' , but it is less stable in period 2.
\end{itemize}

These findings can be confronted with the economic policies that characterize these two periods. Since the 90s a large number of publications dealing with the labor market, were devoted to the theme of flexibility as the preferred means of return to full employment and growth without inflation. This is particularly highlighted by the distinction between job stability, throughout the year, and employment flexibility: the worker is busy most of the year but through changes between successive jobs short in duration. What we see with the Table \ref{tab:transitionperiod} is a reduction of the probability of maintaining a stable and well-paid employment in average (class 5), while the probability of remaining unemployed (class 1) or in precarious employment (class 2) has increased slightly in the period 2. One may say that the new form of employment named \textit{flexibility} seems to have very weak effects, except for the reduction of the stability of  Class 5.\\

It is interesting to study these transitions for women, in order to see if these global results are verified in a gender perspective. See Table \ref{tab:transitionperiodwomen}.\\

\begin{table}[h]
\begin{center}
\begin{tabular}{l|rrrrr}
Period 1 &	Class 1 &	Class 2 	&	Class 3  &	Class 4 	&Class 5  \\
1984-92 	&	2058	&	2599	&	692	&	594	& 1847		\\
\hline
Class 1	&	\textbf{62.80}	&	19.38		&	13.57	&	2.64		&	1.62	\\
Class 2	&	11.56	&	\textbf{60.38}&	6.43	&	6.86	&	14.77	\\
Class 3	&	31.41	&	\textbf{40.90}	&	19.70	&	6.46	&	1.54	\\
Class 4	&	4.55	&	32.71	&	5.64	&	\textbf{41.94}	&	15.16	\\
Class 5	&	4.41	&	7.24	&	2.75	&	4.81	&	\textbf{80.00}	\\
\hline
Period 2	&	Class 1  &	Class 2  &	Class 3 &	Class 4 &	Class 5  \\
1993-2003	&	1270	&	3230	&	355	&	7963	&	2694	\\
\hline
Class 1	&	\textbf{60.30}	&	23.87	&	9.99	&	3.16		&	2.68	\\
Class 2	&	7.46	&	\textbf{65.83}	&	3.55	&	8.18		&	14.98	\\
Class 3	&	20.62	&	\textbf{51.16}&	13.91	&	7.97		&	6.33	\\
Class 4	&	3.69	&	33.95	&	2.07	&	\textbf{41.87}	&	18.43	\\
Class 5	&	3.11	&	16.11	&	1.36	&	5.88	&	\textbf{73.53}	\\
\end{tabular}
\end{center}
\caption{\label{tab:transitionperiodwomen} Transition matrices for women, period 1 and period 2,  the figures below the class names are the class sizes, other values are expressed as percentages, values in bold are maxima.}
\end{table}

We can observe that women stay  longer  in Class 2 (that is the precarious jobs class) in period 2 than in period 1. Conversely, as the mean is almost constant, it means that  men more often leave  precarious jobs to get Class 5 (goods jobs).

At the same time, in period 2, women are less often in Class 5 (goods jobs). When they are initially out of the market (Class 3), they stay shorter in class 3 (withdrawal) and they change more often towards Class 2 and less often towards to class 1 (unemployment).\\

For each period, we can compare the observed distributions of individuals across the five macro-classes (average for 4 transitions over the first period and 5 transitions over the second one)  to the theoretical limit distributions, computed under the hypothesis that everything in the environment stays unchanged during the period. \\

The limit distribution is estimated by iterating the transition matrix. As shown by Markov chain theory (\cite{hoel1987} or \cite{billing1979}), the powers of the transition matrix converge \footnote{when the Markov chain is irreducible, i.e. when all the probabilities to go  from $i$ to $j$, in every number of steps, are strictly positive}, to a matrix where all rows are equal to the limit distribution. So this limit distribution does not depend anymore on the starting value. \\

The empirical and theoretical distributions are displayed  in Table \ref{tab:limit}.
We see  that there is a change between periods 1 and 2. The theoretical and observed distributions are closer in period 2 than in period 1. This indicates that the system has become more stable, i.e. the successive distributions are approximately the same during period 2. 

\begin{table}[htpb]
\begin{center}
\begin{tabular}{p{7cm}|p{1cm}|p{1cm}|p{1cm}|p{1cm}|p{1cm}}
     &\textbf{C1}  & \textbf{C2} & \textbf{C3}  & \textbf{C4}  & \textbf{C5} \\ 
\hline
Empirical distribution (first period) 	&	0.19	&	0.37	&	0.05	&	0.12	&	\textbf{0.27}	\\
Limit distribution (first period) 		&	0.14	&	0.33	&	0.04	&	0.12	&	\textbf{0.37}	\\
\hline
Empirical distribution (second period)	&	0.11	&	0.38	&	0.03	&	0.12	&	\textbf{0.36}	\\
Limit distribution (second period) 		&	0.08	&	0.34	&	0.02	&	0.13	&	\textbf{0.42}	\\
\end{tabular}
\end{center}
\caption{\label{tab:limit}  Empirical and limit distributions, period 1 and 2.}
\end{table}

We observe that the previsions are quite good for classes 1 to 4, but that there is a large discrepancy for class 5. That fact may suggest that the hypothesis of the stationarity during each period is not totally founded. However we see that the drifts are in the same direction.

\section{Discussion and conclusion}

\subsection{Results on SOM alternative topologies}

This paper has investigated several alternative topologies to the classical sheets used with the SOM algorithm. To do so, the SOM algorithm was slightly generalized to deals with arbitrary topology defined by un-oriented graphs. In this setting, two particulars topologies where advocated since they present natural interest in terms of analysis and interpretations, one based on a star shaped graph called SOS and a second one based on distinct strings called "D-SOM". These two topologies were compared with a classical one on an economical dataset and the "D-SOM" was found to be relevant and to ease the interpretation of the produced projection and quantization. Furthermore, the two levels of analysis offered by these two topologies that we call the macro class level and the class level was also interesting for interpreting the projection in our case study. They are however remaining points to be tackled, in particular with respect to topology selection. This difficult problem, was addressed in this paper using prior knowledge from the field study and by the calculation of projection quality indicators based quantization error. But such indicators suffer from several disadvantages, they are in particular sensitive to the elasticity of the topology \textit{i.e.} a topology with less edges will be favoured by such indicators. This may be an interesting path for further works together with the analysis of new type of topology.

\subsection{Discussion about some recent publications on the issue of the functioning of the real labor market}

Some recent publications deal with the labor market functioning: see for example \cite{fre05}, \cite{dav08}, \cite{forni04}, \cite{kri02}, \cite{pri11}, \cite{dan05}, \cite{euro13}. 
The main objection  we can do to these publications is methodological: they use a  macroeconomic  approach to explain a behavior of the individual present on a single market and following a specified dynamics and they ignore difference due to gender. An accurate treatment of this behavior cannot be obtained through macroeconomic studies, while the authors want to check a real heterogeneity in the labor market. This is the case, for instance, in studies initiated by the European Union on the quality of employment, in the spirit of the Laeken indicators (\cite{fre05}, \cite{dav08}). \\

The reduction of individual information by aggregation procedures and the use of specific statistical treatments (Common Factor Model, \cite{forni04}) lead to results as the so-called "trend improvement in quality of employment" like in most institutional publications of the European Union. This appears to be in complete contradiction with most of real labor markets, in other words the idea of quality has been reduced to a single dimension ("to be occupied over the whole year" for instance) instead of a multidimensional concept. In real situations, the quality of a job cannot be enhanced over all the dimensions simultaneously, as we observed for the PSID data.\\

The irrelevance is much worse if we refer to the works based on the notion of a representative agent \cite{blanchard1992} (chapter 2, pp. 37-90), where all individuals are supposed to behave like a typical agent. Thus excluding by construction the heterogeneity (see for example some recent works as \cite{kri02}, \cite{pri11}, \cite{dan05}).\\

At the opposite of the last ones, interesting results are obtained in studies conducted at a microeconomic level (\cite{euro13}) to highlight the question of quality of jobs and its evolution over time.\\

\subsection{Conclusion: Strong results obtained with the classification}

The results can be assessed in three perspectives. 
The first one is the question of the heterogeneity of the actual labor market. Our approach challenges the standard paradigm by providing particularly clear results on a real labor market. In fact what is called the labor market is a set of components of very different qualities in the sense that they can be completely opposed in some dimensions. This is very clear using simple statistic indicators applied to the eight quantitative variables used by the algorithm. The five main classes are well contrasted by means of two or more of these variables. The titles given to these classes are a practical expression of the obtained multidimensional differentiation . As we say before, this differentiation cannot be represented by some unique  scaled variable (score). These components are the permanent structure of the market even if discrepancies may occur over the time, each component having its own dynamics.\\

Another perspective is that of the dynamics of individual situations as it can be represented using the panel structure of the data. This is obtained with the transition matrices built with the classes. For each individual, classification provides the class  he belongs to for each observed year.\\

The observation of empirical probabilities of transition between classes presented in table \ref{tab:transitionperiod} shows that the change possibility strongly depends on the present situation:
\begin{itemize}
\item Firstly the probability of remaining in the class of "good jobs" is 4 out of 5 chances in the first period and 3 out of 4 chances for the second one
\item Secondly, when someone is in a class of low quality jobs he/she has 2 chances out of 3 to stay in it against 1 chance out of 6 or 7 to move to the class of good jobs. This is true for both periods.
\end{itemize}

This analysis is not the explanatory model of the job market: this one remains to be built. It is simply a step in this construction.
Contrary to the idea often put forward, the transfer of the unemployed to new forms of precarious employment (also called "flexibility") is not clearly observed. The most visible movements between classes, considering the whole sample, do not verify this assertion and unemployment remains a relatively stable situation, in the second period as in the first one. In the second period there seems to be a much less frequent move back to unemployment after a temporary withdrawal from the labor market.\\

The last perspective is that of gender. 
The share of those who remain in a status of low activity (activity very restricted or unemployment, Class 1) is very stable and represents a mean of more than 56  \%, for both genders, see Tables \ref{tab:transitionperiod}, \ref{tab:transitionperiodwomen}.\\

But it is not the same for those who are in good jobs (Class 5), see Tables  \ref{tab:transitionperiod},  \ref{tab:transitionperiodwomen}. 
In fact, the overall results mask the real evolution of women’s situation on the whole period. For the complete sample (men and women), the probability to remain in good jobs decreases  slightly  (from 81 \% to 76 \%) from the first period to the second. 
In fact this average  is misleading because simultaneously the probability remains equal   for men (about 82\%) and decreases for women (from 80   \% to 73 \%). This indicates a deepening of the gap between men and women in terms of maintaining into the category of good jobs.\\

In terms of wages, the results are less clear given the arbitration, already mentioned, between job tenure and the level of wages. 
For high values of the seniority, the level of remuneration for men is significantly higher than that of women (about 50 \%), both in the first period and the second period.
For low values of seniority (high levels of wages) the difference is in favor of men but in a much smaller proportion.\\

\subsection{Future work}

As we mention previously, unfortunately, in PSID survey, poor quality of data on qualifications and location of employment in terms of branches of industry (very imprecise classifications, lack of updated information in each survey wave) does not allow to specify the content of classes in these areas. To overcome these drawbacks, we intend to repeat this classification technique and the exploitation of results with another large database, the European Community Household Panel (SILC) built over 27 countries, and now available (on demand) on line on the Eurostat website \footnote{\url{http://epp.eurostat.ec.europa.eu/portal/page/portal/statistics/search_database}}.


Moreover, according to  some recent studies  \cite{bartolucci2009}, we plan to improve the theoretical study of the heterogeneous labor market,  by modeling the 
transitions between segments over time using some econometric tools, like a dynamic logit model, with measured factors and controlling 
unobserved  heterogeneity in a   Hidden Markov Chain modelization.


\bibliography{wsom2012}

\end{document}